\documentclass[sn-vancouver,Numbered]{sn-jnl}

\usepackage[numbers]{natbib}
\usepackage{graphicx}%
\usepackage{afterpage} 
\usepackage{multirow}%
\usepackage{amsmath,amssymb,amsfonts}%
\usepackage{amsthm}%
\usepackage{mathrsfs}%
\usepackage[title]{appendix}%
\usepackage{xcolor}%
\usepackage{textcomp}%
\usepackage{manyfoot}%
\usepackage{booktabs}%
\usepackage{algorithm}%
\usepackage{algorithmicx}%
\usepackage{algpseudocode}%
\usepackage{enumitem}
\usepackage{xcolor}\usepackage{listings}%

\usepackage{soul}

\usepackage[english]{babel}
\usepackage[autostyle, english=american]{csquotes}
\MakeOuterQuote{"}

\usepackage{geometry}

\usepackage{caption}
\usepackage{subcaption}

\newcounter{example}[section]
\newenvironment{Step}[1][]{\refstepcounter{example}\par\medskip
   \noindent \textbf{Step~\theexample. #1} \rmfamily}{\medskip}

\begin{document}

\title[Article Title]{A solution for co-locating 2D histology images in 3D for histology-to-CT and MR image registration: closing the loop for bone sarcoma treatment planning}


\author*[1]{\fnm{Robert} \sur{Phillips}}\email{robertphillips8023@gmail.com} 

\author[2]{\fnm{Constantine} \sur{Zakkaroff}}
\nomail
\author[4]{\fnm{Keren}
\sur{Dittmer}}\email{}
\nomail

\author[5]{\fnm{Nicholas}
\sur{Robillard}}\email{}
\nomail

\author[6]{\fnm{Kenzie}
\sur{Baer}}\email{}
\nomail

\author[1,3]{\fnm{Anthony} \sur{Butler}}
\nomail

\affil[1]{\orgname{The University of Otago - Canterbury},\orgdiv{Pathology and Biological Sciences}, \orgaddress{\city{Christchurch}, \country{New Zealand}}}

\affil[2]{\orgdiv{UC Business School}, \orgname{University of Canterbury}, \orgaddress{\city{Christchurch}, \country{New Zealand}}}

\affil[3]{\orgname{Te Whatu Ora},\orgdiv{Radiology},  \orgaddress{\city{Christchurch}, \country{New Zealand}}}

\affil[4]{\orgname{Massey University}, \orgdiv{Tāwharau Ora – School of Veterinary Science}, \orgaddress{\city{Palmerston North}, \country{New Zealand}}}

\affil[5]{\orgname{Te Whatu Ora (Canterbury Health Labs)}, \orgdiv{Anatomic Pathology}, \orgaddress{\street{Riccarton Ave}, \city{Christchurch}, \country{New Zealand}}}

\affil[6]{\orgname{Ossis Limited}, \orgaddress{\city{Christchurch}, \country{New Zealand}}}










\abstract{
This work presents a proof-of-concept solution designed to improve the accuracy of radiographic feature characterisation in pre-surgical CT/MR volumes.
The solution involves 3D co-location of 2D digital histology slides within \textit{ex-vivo}, tumour tissue CT volumes.

In the initial step, laboratory measurements obtained during histology dissection were used to seed the placement of the individual histology slices in corresponding tumour tissue CT volumes. The process was completed by aligning corresponding bone in histology images to bone in the CT using in-plane point-based registration.

Six bisected canine humerus datasets of \textit{ex-vivo} CT and corresponding measurements were used to validate dissection placements. Digital seeding exhibited a plane misalignment of 0.19 $\pm$ 1.8\,mm. User input sensitivity caused 0.08 $\pm$ 0.2\,mm in plane translation and between 0 and 1.6\textdegree ~deviation. These are of similar magnitude to reported misalignment of 0.9-1.3\,mm and 1.1-1.9\textdegree{} in related prostate histology co-location~\cite{Gibson2013quantifyTissueDeformationInProstate}.

Although this work only reported on animal femur sarcoma CT images and histology slide images, the solution can be generalised to various sarcoma geometries and presentation sites. Furthermore, the solution co-locates high-fidelity data to advance tissue characterisation with minimal disruption to existing
clinical workflows. Improved tissue characterisation accuracy, supported by the resolution of histology images, can enhance surgical planning accuracy and patient outcomes by bringing the insights offered by histology closer to the start of the presentation-diagnosis-planning-surgery-recovery loop.

}




\maketitle

\pagebreak

\begin{figure*}[t]
    \centering
    \includegraphics[width=0.9\textwidth]{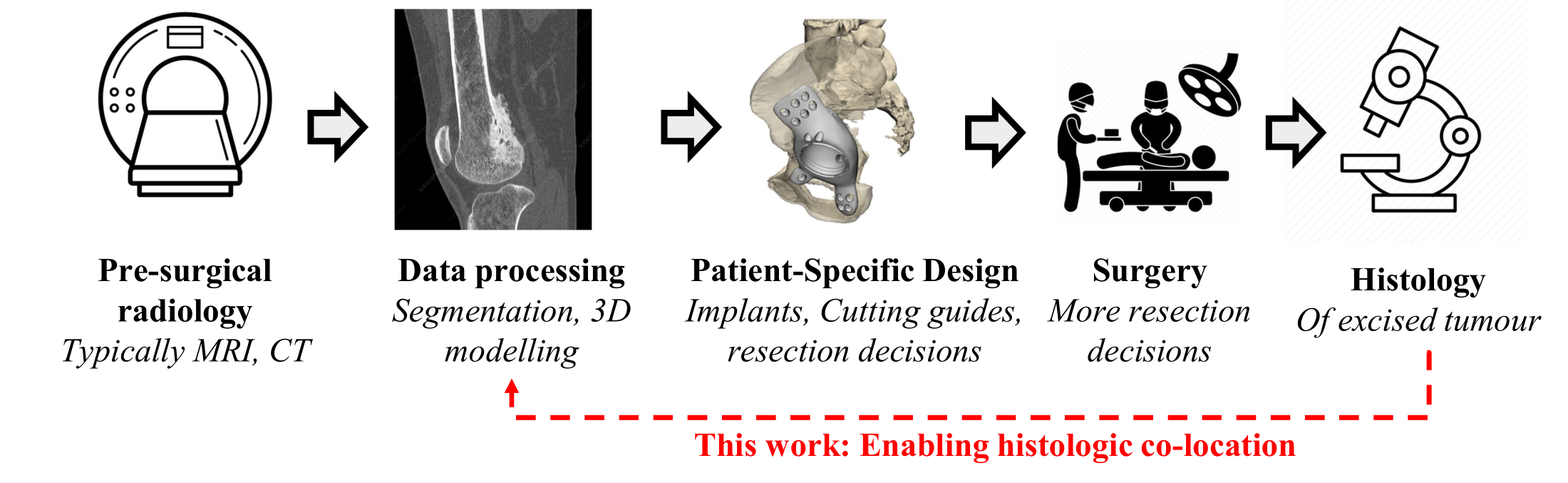}
    \caption{Reference patient-specific clinical workflow for bone sarcoma.}
    \label{Fig:SarcomaWorkflow}
\end{figure*}


\section{Introduction}

We present a computational and tissue processing solution for the co-location of 2D histology images and 3D sarcoma radiology volumes. The solution uses two CT scans of each half of a bisected tumour specimen following surgery and incorporates the taking of offset measurements in laboratory processing. We also outline a sequence of operations for further registration with pre-surgical CT and MR. 
The combination of radiologic datasets with co-located histology images and labelmaps will support image interpretation and segmentation confidence when making subsequent surgical resection margin decisions.
Adopting this solution in clinic will enable the co-location of high-fidelity training and validation data to support segmentation and surgical technology.

Histology slides, radiology and an improved laboratory workflow from previous experimental sarcoma work~\cite{Phillips2023} were used to develop and refine computations. An additional canine sarcoma dataset, collected following routine veterinary surgery, demonstrates continuity of the overall histology image co-location solution.

\subsection{Challenges and Considerations in the Treatment of Bone Sarcomas}
Bone sarcomas are some of the most challenging tumours to treat~\cite{bosma2018cadaveric, parry2016osteosarcoma}. Treatment and management of these tumours is less well understood because they are rare and heterogeneous, consisting of a variety of cell types, sub-populations, and tissue substructures~\cite{evrard2019resection}. They are over-represented in younger populations and, if not fatal, often impact mobility and quality of life~\cite{Ballantine2020, parry2016osteosarcoma}. Orthopaedic surgery is indicated for all curative treatment along with chemo radiation in most cases~\cite{Grimer2010, gerrand2016behalf}. Complex procedures, such as hemipelvectomy, may be necessary when in the pelvis or other complex anatomy~\cite{BoneSarcomasGuidelines}. Despite inclusion of indicated multi-agent systemic chemotherapy, the reported five-year overall survival rates for the most common bone sarcoma have not improved much beyond 27 percent here~\cite{parry2016osteosarcoma,Ozaki2003OSpelvis}. 


Resection margins aim to ensure full tumour removal. However, healthy tissue loss with these margins impacts patient outcomes. The recommended curative treatment is a \emph{wide} resection, which aims to include 10\,mm or 20\,mm of healthy tissue around the removed tumour~\cite{Grimer2010, Biermann2013bone}. Sometimes wider margins are taken due to factors like the patient's risk profile, sarcoma sub-type or geometry of the surgical cuts. Unfortunately, damage to muscles, nerves and blood vessels resulting from healthy tissue removal impair post-surgical mobility and body function. These complications can burden the patient, family, community, and the health system~\cite{Shafrin2017MobilityBurden}. 

Clinical teams face difficult decisions regarding resection margins. They must remove functional healthy tissue to ensure surgical success while also attempting to minimise harm from loss of the same healthy tissue~\cite{Da2021, mcmaster2001chordoma, Bosma2019nav, walcott2012chordoma}. Unfortunately there is limited data to predict the surgical success of different resection options. This forces clinical teams to make important decisions with minimal quantitative guidance.

Patient-specific technology shows promise for reducing surgical margins. For example, computer-aided design (CAD) models help sarcoma treatment teams \emph{salvage} healthy tissue, especially in areas with complex anatomy~\cite{Boyle2023CustomImplant}. 
However, prospective studies face challenges in evaluating custom implants, cutting guides, surgical navigation against traditional orthopaedic benchmarks due to limited sample availability~\cite{evrard2022quality, park2018bone, bosma2018cadaveric}.

The need for more samples presents a challenge. Traditional orthopaedic oncology research often spans decades before confidently recommending new surgical techniques~\cite{Enneking1980manual, kawaguchi1995new, Grimer2010}. 
However, rapid evolution in salvage technology from the advent of 3D printers and better access to computers highlights the need for a solution to understand the inherent risk of different surgical techniques. It has been proposed that high-fidelity data be used to assess sarcoma treatment~\cite{Cannavo2019qualEvalDeliniation}. Unfortunately, the utilisation of such data is limited from constraints in current post-surgical analysis reporting~\cite{gomez2019assessmentChemo, durr2014resection}. 

Medical image segmentation and 3D modelling (Figure~\ref{Fig:SarcomaWorkflow}) are common techniques used across almost all patient-specific technologies~\cite{VandenBroeck2014}. Using segmented CAD models to guide resection decisions is immensely valuable to clinical teams. However, this process places significant reliance on the engineers and machines performing segmentation, and the radiologists and surgeons who check the CAD model outputs. The increasing reliance on CAD models to determine resection margins introduces a limitation, as tumour segmentation remains subjective and difficult to quantify. 

Some studies have evaluated the radiological presentation of sarcoma in one dimension using the clinical standard of histopathology, the diagnosis of disease by microscopic evaluation~\cite{Gulia2019boneHistReg, malherbe2020assessment, ahmad2014accuracy, Onikul1996boneHistReg}. While these radiologic evaluation techniques are clinically relevant, they only capture a fraction of information contained in histology. Furthermore, no histology registration studies have been found suitable for tissue of the pelvis or other complex, curved anatomy. 
Areas where resection margin assessment could greatly benefit patients. 

Co-location of 2D histology images in 3D radiology volumes has proven to be effective for validating tissue boundaries and tumour margins for soft tissue~\cite{Pichat2018HistRecon}. Many novel techniques have been developed to co-register \textit{ex-vivo} histology and \textit{in-vivo} MR~\cite{reynolds2015prostateRegHist, Gibson2012prostateRegHist, Park2008prostateRegHist}. However, few published techniques are appropriate for bone, with the most notable requiring cryogenically frozen mice~\cite{Seeley2014histBoneRecon}. Cryogenesis is not always practical for larger clinical sarcoma. 

In contrast, this paper outlines a solution to co-locate histology images and register them to medical image volumes for bone sarcoma. The proposed solution accommodates variations in shape, size, and tissue heterogeneity, and can be performed with minimal disruption to most clinical pathology workflows.

\subsubsection{Accuracy of radiologic segmentation}
Clinical radiology is widely used in innovative salvage techniques~\cite{durr2014resection}. CAD models are increasingly employed to determine resection margins before surgery, often fixed in printed devices, tools, or digital surgical navigation systems~\cite{Boyle2023CustomImplant, cho2017augmented}. 
Nevertheless, a \emph{critical assessment} of resection margins and therapy techniques is needed to realise the full opportunities from advanced surgical technology~\cite{durr2014resection}, of which segmentation is an integral part.

Segmentations of sarcoma can vary due to scan resolution, imaging artefacts like partial volume averaging, inhomogeneity, or motion and different radiologic interpretations. While medical image resolution and radiology protocols for diagnosis and segmentation are continually evolving~\cite{Guedes2021clinicalImageEval}, their interpretation remains fundamentally subjective. Technology providers are developing artefact reduction technologies for CT and MR, though these come with their own challenges. Other recent research focuses on automating model segmentation to reduce human variation and support clinical translation~\cite{MONAI}. However, all these initiatives face the same issue: it is difficult to quantify how well individual image features approximate physical tissue boundaries~\cite{Cannavo2019qualEvalDeliniation}.

Co-location of 2D histology and 3D radiology image features is valuable for quantifying image segmentation and radiomics as it links microscopic tissue properties to macroscopic information~\cite{Pichat2018HistRecon}. Our solution supports histology image co-location within radiology volumes, facilitating quantitative and semi-quantitative image processing for bone sarcoma.

\subsection{Histology registration current state-of-the-art}
Histopathology is the clinical standard for tumor diagnosis and is used to guide post-surgical treatment~\cite{Mbalaviele2020BoneBiologyBook, Bishop2024endToendWorkflow}. It employs histology techniques to assess tumour boundaries. Despite its relatively low cost, histology provides a substantial amount of high-fidelity data and is expected to remain a cornerstone of pathology~\cite{Rosai2007histologyWillRemainCnrstnPathology, Bishop2024endToendWorkflow}. 

Histology techniques involve dissecting resected specimens, fixing them to prevent autolysis, and performing several immersion steps. Thin shavings from each dissection are then stained on glass slides and analysed under a microscope. 

Histology literature describes how slides should quantitatively be measured and interpreted~\cite{2024RCPApathologycutup}. There are a number of classification systems which describe what tissue features and margins mean for post-surgical tumour treatment. The Residual (or \textit{R}) classification in the Tumour, Node, Metastasis (TNM) system is a common metric used to check surgical success~\cite{gomez2019assessmentChemo, evrard2022quality}. 

Increasing digitisation provides very high-fidelity images for use in treatment but also processing beyond clinical workflows~\cite{Rosai2007histologyWillRemainCnrstnPathology, Mbalaviele2020BoneBiologyBook}. Over the last three decades, successful co-location of microscopic histology and registration with macroscopic radiology volumes have been performed~\cite{Pichat2018HistRecon}. Histology can produce a discontinuous stack of images with an effective pixel size of 5\,$\mu m$ by 5\,$\mu m$ for 4\,$\mu m$ thick slices of tissue, creating a volume with tiny voxels if shaved in parallel~\cite{Rusu2020prostateRegHist}. In contrast, MR has a typical voxel size of 0.4 x 0.4 x 4.0\,$mm^3$-\cite{Rusu2020prostateRegHist}.

Unfortunately spatially localising human bone sarcoma histology to MR remains unattainable with current solutions~\cite{Phillips2023}. 

This paper employs a novel pathology processing approach to assist histology image co-location within bone sarcoma radiology~\cite{Phillips2023}. It can stand alone, or be added to the start of existing open source histology image co-location solutions~\cite{Rusu2020prostateRegHist}. 

\subsection{Anticipating clinical use early}\label{Bacground_subsection_Clinical_relevance}
Registration between \textit{in-vivo} imaging modalities, such as CT and MR, is well established. And augmentation with histology can make high-fidelity information available at a low cost~\cite{Pichat2018HistRecon}. However, accurately co-locating 2D histology images in a 3D context within established clinical environments is challenging~\cite{Bishop2024endToendWorkflow}. 
Prior research has highlighted three key requirements of any histology co-location solution: 1) to permit specimen slicing and dissecting ('cut-up') in line with pathology protocols; 2) to provide co-location accuracy that is robust to the variation of tissue appearance on radiology and histology; and 3) to provide a quantitative evaluation of co-location error~\cite{Gibson2012prostateRegHist}. 

The collection of high-quality 3D bone histology datasets should prioritise the simple adaptation of existing pathology protocols before overhauling entire systems or requiring specialised equipment. It is noted that a) pathology protocols vary due to numerous factors, such as equipment availability and staffing levels. b) bone sarcoma is highly variable in shape and size. c) bone sarcoma is rare, making it crucial for the solution to be usable in as many sarcoma centers as possible. Minimising protocol change enhances resilience by leveraging the evolved knowledge and expertise within institutions.

While previous studies have explored histology co-location in the context of the prostate, this work focuses on bone anatomy and integration with the reference sarcoma workflow (Figure~\ref{Fig:SarcomaWorkflow}). Large bone specimens are slow to fix and require prior dissection to ensure proper penetration of fixative agents (e.g. formalin)~\cite{Mbalaviele2020BoneBiologyBook, 2024RCPApathologycutup, PathReportDissectionPrior}. Bone specimens need further demineralisation prior to microtoming and slide preparation. These practical limitations are believed to contribute to the lack of histology co-location work with clinical bone sarcoma.

\textit{Nonparallel} tissue dissection is more relevant to clinical workflows than parallel~\cite{Gibson2012prostateRegHist, Gibson2016prostateRegHist}, especially for sarcoma. Our solution uses the bisection identified on \textit{ex-vivo} CT to introduce a reference surface, restoring the 3D spatial orientation of \textit{nonparallel} 2D digitised histology slides, and facilitating further image volume registration.

Any histology co-location solution should collect data to support both soft tissue and bone segmentation. Emerging orthopaedic surgical technology aims to minimise damage to healthy soft tissue~\cite{bosma2018cadaveric}. Confidence in image characterisation will be immediately useful in making resection decisions around nerves and blood vessels.

Previous work found the presence of both bone and soft tissue in bone sarcoma specimens to be problematic~\cite{Phillips2023}. However, the rigidity of bone has been utilised to maintain tissue structure during cut-up without need for external apparatus. Additionally, bone’s lower level of shrinkage compared to other tissues~\cite{Cannavo2019qualEvalDeliniation} presents an opportunity to support robust co-location.

Clinical adoption of methods for prostate histology co-location have been slow for challenges in physical processing, lack of easy access methods for data combination, and time constraints in clinical pipelines~\cite{Rusu2020prostateRegHist}. Prostate cancer groups first shared their methodology and solutions for histology co-location~\cite{reynolds2015prostateRegHist, Rusu2017method, Gibson2012prostateRegHist} which informed subsequent clinical data collection~\cite{reynolds2019prostateRegHist, Rusu2020prostateRegHist, Gibson2016prostateRegHist}. Their solutions have continued to evolve and grow in recent years~\cite{Shao2021prostateRegHistML}. 
Our work builds on the most promising methodology for bone sarcoma by \citeauthor{Phillips2023}. We introduce a novel computational solution, supported by proof-of-concept \textit{ex-vivo} animal data.

\clearpage
\begin{figure*}[tbp]
    \centering
    \includegraphics[width=\textwidth]{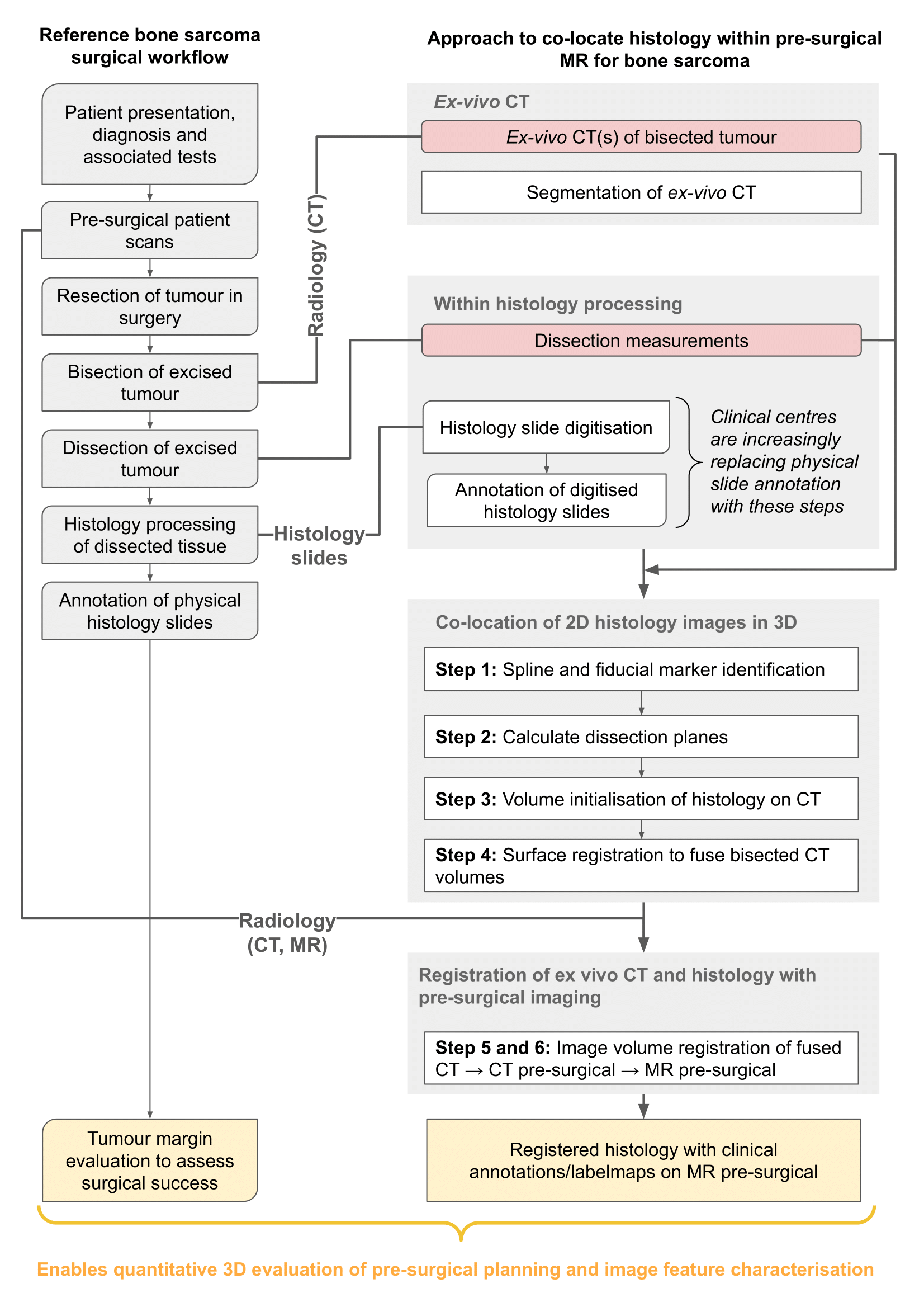}
    \caption{Image left is a reference sarcoma surgical workflow, while image right is an outline of the image co-location and registration steps this work presents. Lines between are where information is pulled from the surgical workflow. Red, rounded shapes are additional operations performed on tissue in pathology processing. 
    }
    \label{fig:ClinicalWorkflow}
\end{figure*}

\clearpage

\section{Method}
Extended consultation with surgeons, radiologists and pathologists was a focus of previous work~\cite{Phillips2023}. This identified the requirements and constraints of their sarcoma workflows.
See Figure~\ref{fig:ClinicalWorkflow} for an overview of how the solution would fit into the reference surgical workflow.

\subsection{Steps in solution computations}
\subsubsection{Definitions}
Note the term \textit{'bisection'} refers to approximately half a tumour specimen following the initial bandsaw cut but before \textit{ex-vivo} CT. The term \textit{\textquoteleft dissection\textquoteright} refers to tissue from subsequent bandsaw cuts of the two bisections after \textit{ex-vivo} CT. 

More definitions:
\begin{description}[labelwidth=2cm,leftmargin=!]
    \item[$CT_{post,1 \& 2}$] CT of bisected volumes 
    \item[$H_{i}$ ] Digitised histology image for $i^{th}$ slide with associated labelmaps 
    \item[$d_{a,i}$ \& $d_{b,i}$ ] Laboratory measurements from fiducial reference ($f_{ref}$) to the two corners made by the bisection surface and each dissection cut 
    \item[$P_{i}$ ] Assigned planes for $i^{th}$ dissection cuts (on which $H_i$ are initialised)
    \item[$H_{vol}$ ] Histolgoy volume, featuring a sparse 3D surface of bone annotations
    \item[$T_{fuse}$ ] The transform to combine \textit{ex-vivo} bisection CT volumes into $CT_{post, fused}$ 
    \item[$CT_{pre}$ ] Pre-surgical CT volume 
    \item[$MR_{pre}$ ] Pre-surgical MR volume 
\end{description}

\subsubsection{Pre-processing}\label{Methods:StepsToSolution}
Bone segmentation in each $CT_{post}$ volume began with a threshold at 226 HU (Hounsfield Unit)~\cite{VandenBroeck2014} (default in some Materialise Mimics\textsuperscript{\copyright} versions), followed by manual adjustment in 3DSlicer~\cite{3DSlicerFedorov2012}. Segmentations and digital histology binary labelmaps were fed into a computational process outlined below. Labelmaps were from QuPath~\cite{QuPathBankhead2017}.

\begin{Step}\newline
    \textit{Input:} Bisected bone segmentation for each $CT_{post}$ volume.\newline
    \textit{Output:} The bisected bandsaw surface was contoured by two markup splines and a markup fiducial reference.\newline 
    \textit{Process:} Two 3DSlicer
    markups ($spline_{ma}$ and $spline_{mb}$) were manually placed along bone~/fibrous tissue edges of the bisection as seen in the segmentation, with an additional markup point at $f_{ref}$. These were exported from 3DSlicer into separate files.
\end{Step}

\begin{Step}\newline
    \textit{Input:} The Cartesian position of $spline_{ma}$, $spline_{mb}$ and $f_{ref}$ control points.\newline
    \textit{Output:} Planes assigned ($P_i$) on splines.\newline
    \textit{Process:} Parameterised cubic polynomials were fit to $spline_{ma}$ and $spline_{mb}$ control points with coefficients stored. 
    Each measurement $d_{a,i}$ and $d_{b,i}$ were treated as the Euclidean distance between $f_{ref}$ and the intersection of the $P_i$ with the cubic polynomials.

    The Euclidean calculation was numerically solved through fsolve of Scipy~\cite{Virtanen2020scipy} using a variation on the Newton Raphson method to find the intersect in 3D of $P_i$ with the two polynomials.
    
    At the resulting pair of intersections, parametric polynomials were differentiated. An average normal was calculated. This, combined with the vector between intersections, gave the cross product to define the geometry of dissection cuts as perpendicular to the curved bisection. This process models each approximate bandsaw dissection cut. 

\end{Step}

\begin{Step}\newline
    \textit{Input:} $P_{i}$, each $CT_{post}$ segmentation and $H_{i}$.\newline 
    \textit{Output:} Transforms to initialise $H_{i}$ into $H_{vol}$.\newline
    \textit{Process:} This step consisted of two parts. $P_{i}$ and $H_{i}$ were imported into into 3DSlicer. $H_{i}$ were rigidly transformed from their global import position to $P_i$ using the VTK GetObjectToWorldMatrix (3DSlicer terminal). From here an additional transform scaled and translated binary histology labelmaps in-plane using 3 pairs of landmark points to overlay $H_{i}$ on the respective bisection CT.
\end{Step}

\begin{Step}\newline
    \textit{Input:} Both $CT_{post}$ volumes and bone segmentations. \newline
    \textit{Output:} $CT_{post, fused}$ volume.\newline
    \textit{Process:} The two bisected surfaces from step 1 segmentation were aligned using point-to-point fiducial registration (3DSlicer). Then multiple spherical surfaces points from the \textit{dynamic modeller} module (3DSlicer) were used to refine the registration of the moving surface to the fixed model using \textit{model registration} under the IGT module~\cite{Ungi2016IGT} (3DSlicer). This sequence of registrations combined to give $T_{fuse}$ for aligning bisected CT volumes to output $CT_{post, fused}$ once run through the \textit{stitch volumes} extension (3DSlicer).

\end{Step}

\begin{Step}\newline
    \textit{Input:} $CT_{post, fused}$ and $CT_{pre}$. \newline
    \textit{Output:} Transform to register $CT_{post, fused}$, along with initialised $H_{vol}$, to $CT_{pre}$. \newline
    \textit{Process:} The 'generic (all)' preset was used in Elastix~\cite{3DSlicerFedorov2012} for combining canine volumes. 
\end{Step}

\begin{Step}\newline
    \textit{Input:} $CT_{pre}$ and $MR_{pre}$. \newline
    \textit{Output:} Transform to register $MR_{pre}$ to $CT_{pre}$. Alternatively the total transform could be inverted to register CT and associated $H_{vol}$ to MR. \newline
    \textit{Process:} The Elastix~\cite{3DSlicerFedorov2012} module was again used with the same preset for this.
\end{Step}

\subsection{Testing Plane Assignment (of Step 1 and 2)}
\begin{figure*}[tph]
    \begin{subfigure}[t]{0.49\textwidth}
         \centering
         \includegraphics[width=\textwidth]{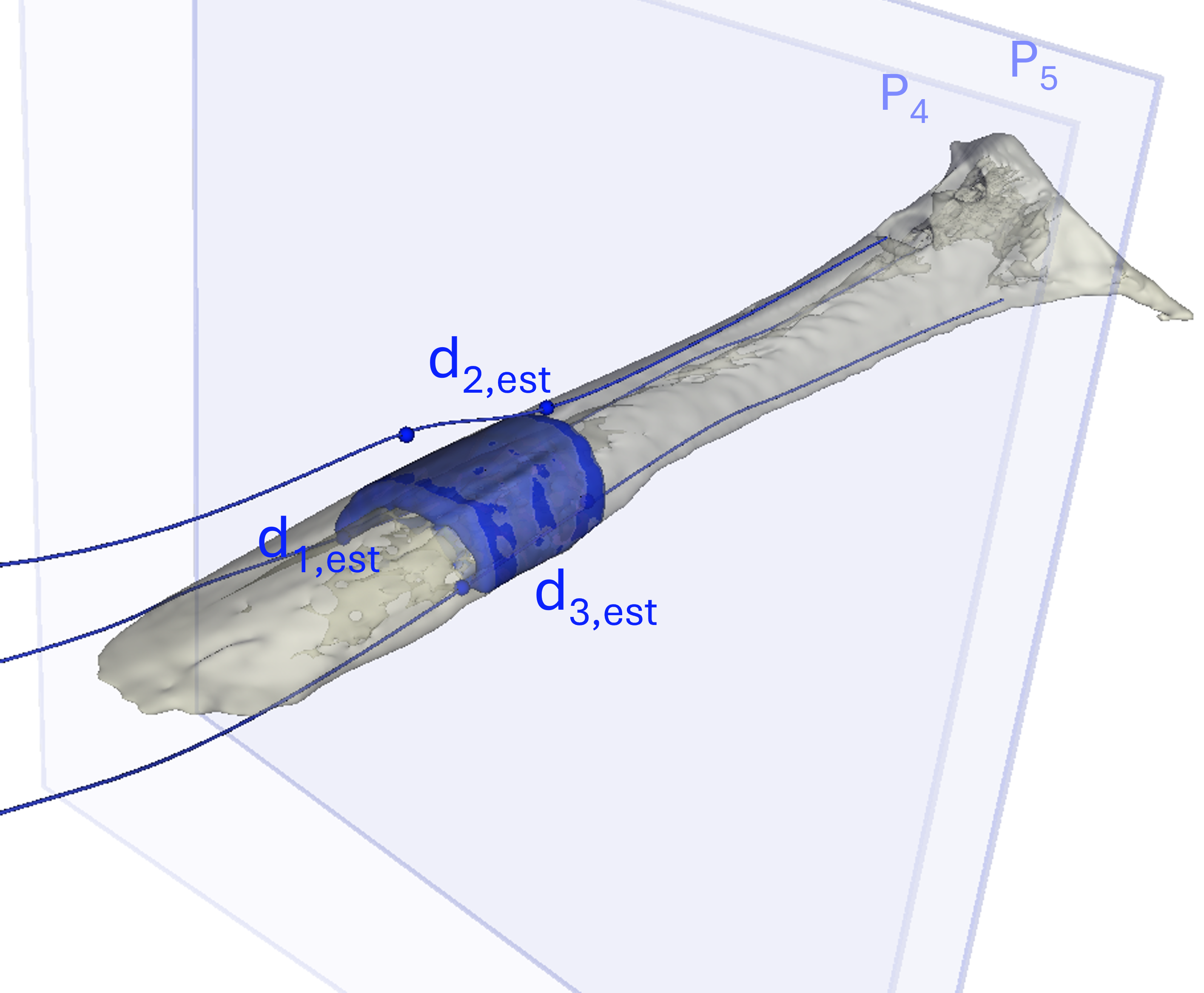}
         \caption{Computed dissection (blue), between two subsequent planes ($P_4$ and $P_5$), overlaid on the bone model.}
         \label{fig:sub1}
    \end{subfigure}
    \hfill
    \begin{subfigure}[t]{0.49\textwidth}
        \centering
        \includegraphics[width=\textwidth]{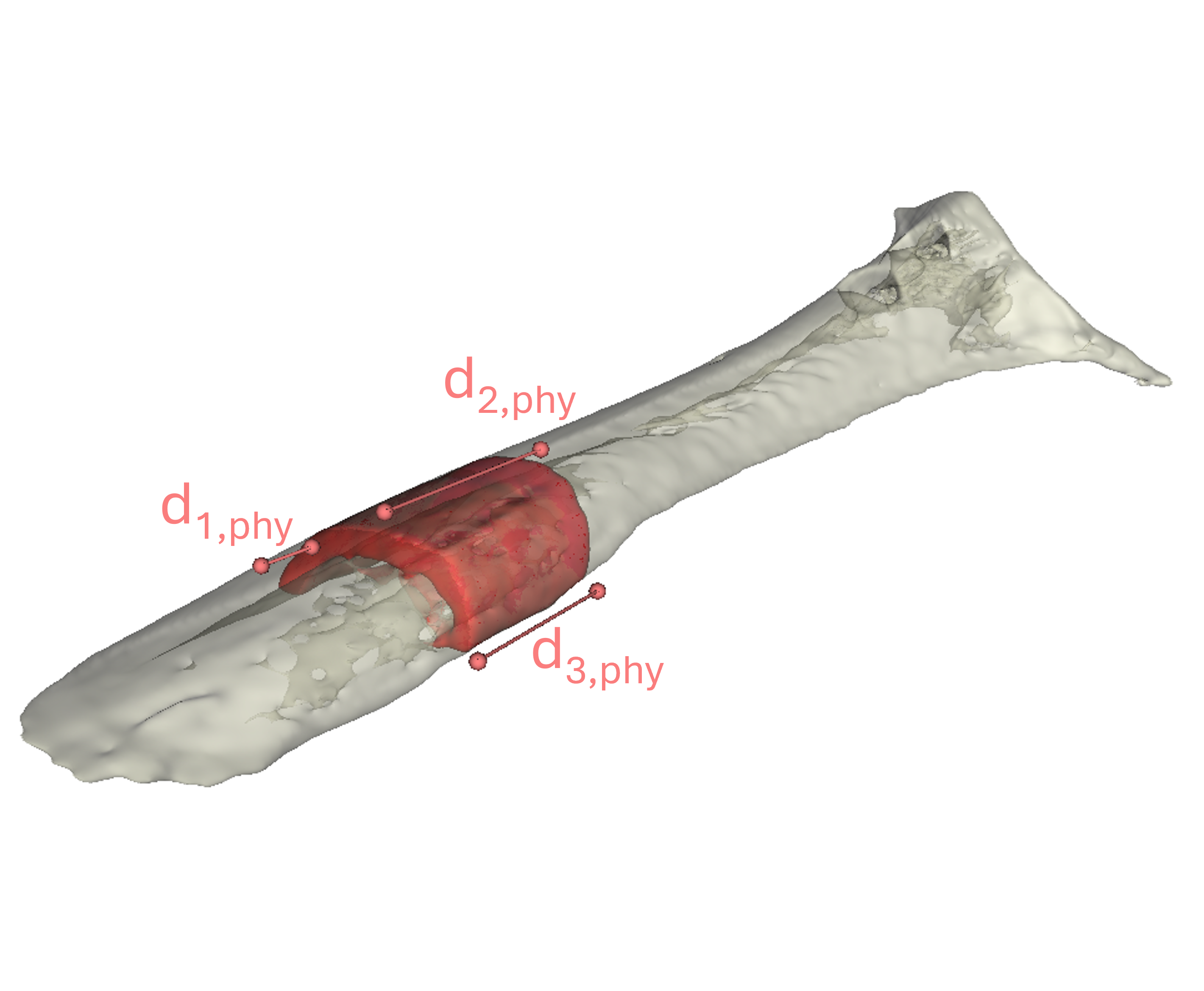}
        \caption{Visualising dissection (red) as handled in the laboratory, overlaid on bone model.}
        \label{fig:sub2}
    \end{subfigure}
    
    \caption{3D bone segmentation models of a whole CT bisection with a single dissection overlaid. The euclidean distances ($d_{1,est}$, $d_{2,est}$ and $d_{3,est}$) from computations were recorded along with physical measurements ($d_{1,phy}$, $d_{2,phy}$ and $d_{3,phy}$) to evaluate plane assignment. Image generated using 3DSlicer.}
    \label{fig:ValidationMain}
\end{figure*}

It is important to appreciate misalignment in histology co-location~\cite{Gibson2013quantifyTissueDeformationInProstate}. 
We tested this by comparing laboratory and computed measurements of dissected tissue size.
Figure~\ref{fig:ValidationMain} illustrates how we collected dissection size measurements, as proxies for the relative positions of dissection cuts, to understand translation and rotation errors in plane assignment.

We paired euclidean distances between original spline intersections on successive planes ($d_{1,est}$ and $d_{3,est}$) with physical vernier caliper measurements ($d_{1,phy}$ and $d_{3,phy}$) from the laboratory.  Similarly, we compared Euclidean distances along a third spline ($d_{2,est}$) that contoured the bone and fibrous tissue surface on $CT$ with a third set of caliper measurements (${d_2,phy}$) for each bisection.

We evaluated plane assignment sensitivity by adjusting the user input. The extremes of spline placement on bone or fibrous tissue in $CT_{post, 1\&2}$ was determined by expert interpretation. We compared planes from calculation of input spline extremes to understand how $P_i$ translation and rotation responded, which indicated the solution’s sensitivity.

\subsection{Proof-of-Concept}
\subsubsection{Data Description}
Clinical animal data used in this work was of amputated canine limbs. Prior owner consent was received for use of data in education and research. Routine veterinary treatment remained unchanged as all imaging and processing was performed post-amputation. This work was granted exception for animal ethics approval because no investigations were performed on, and no treatment plan altered for live dogs.

All images, animal and owner details, were de-identified with DicomCleaner (PixelMed Publishing$^{TM}$) for solution development and testing. Canine specimens were considered a suitable model for humans because canine bone tumours have many clinical similarities to equivalent human disease~\cite{romanucci2023canine}.

\textbf{Experimental Radiology Protocols}\newline 
Multi-parametric MR and CT data was available from previous work to assist solution development~\cite{Phillips2023}. Additional image stacks from a 3T Magnetom Vida (Siemens Healthineers, Germany), Christchurch City, Ōtautahi, New Zealand were used to evaluate the solution in the proof-of-concept. This work used 1\,mm slice sagittal spatial reconstructions highlighting bone (T1) and tumour (T2 Turbo Spin Echo). MR sequence parameters were adjusted for scanning a small FOV. 

High-definition CT of the whole limb ($CT_{pre}$) and bisected specimens ($CT_{post,1}$ \& $CT_{post,2}$) were taken on a Revolution GSI (GE Medical Systems, USA), Lincoln, New Zealand. This work used 0.625\,mm slice axial spatial reconstructions. These were acquired at 120 kVp with a 0.5 pitch and 512 x 512 matrix.

\textbf{Experimental Histology Digitisation}\newline 
Standard-size Haemtoxylin and Eosin (H\&E) stained histology slides were digitised by Awanui Labs in Christchurch City. Large format H\&E slides were digitised at \hbox{12.5 X} zoom with a scanning optical camera setup as no cassettes were present in the country for digitising large format histology on any commercial scanner.

A human clinical pathologist annotated all digital histology images while referencing physical slides under a microscope. However, as high-resolution commercial digitisation systems become more available~\cite{Mazzarini2020Histology}, referencing of physical slides may become less relevant.

\section{Results}

\subsection{Testing Plane Assignment (of Step 1 and 2)}
Our study evaluated the accuracy of plane assignment by comparing computed measurements and physically dissected tissue size. Across 114 measurements on 38 dissections (over 6 bisections), the mean error difference was 0.19 \,mm ($\pm$ 1.8\,mm). Additionally, rotational analysis showed a mean error difference of 0.25 \,mm ($\pm$ 1.9\,mm) for $d_2$ measurements and 0.4 \,mm ($\pm$ 1.6\,mm) for $d_1$ and $d_3$ measurements, based on 38 and 76 measurements, respectively.

Different user input indicated a maximum rotational plane assignment sensitivity of 1.6\textdegree, and mean translation sensitivity of 0.08 $\pm$ 0.2\,mm (n = 114). 

\subsection{Step 3 and beyond}
\begin{figure}[H]
    \centering
    \includegraphics[width=1.0\linewidth]{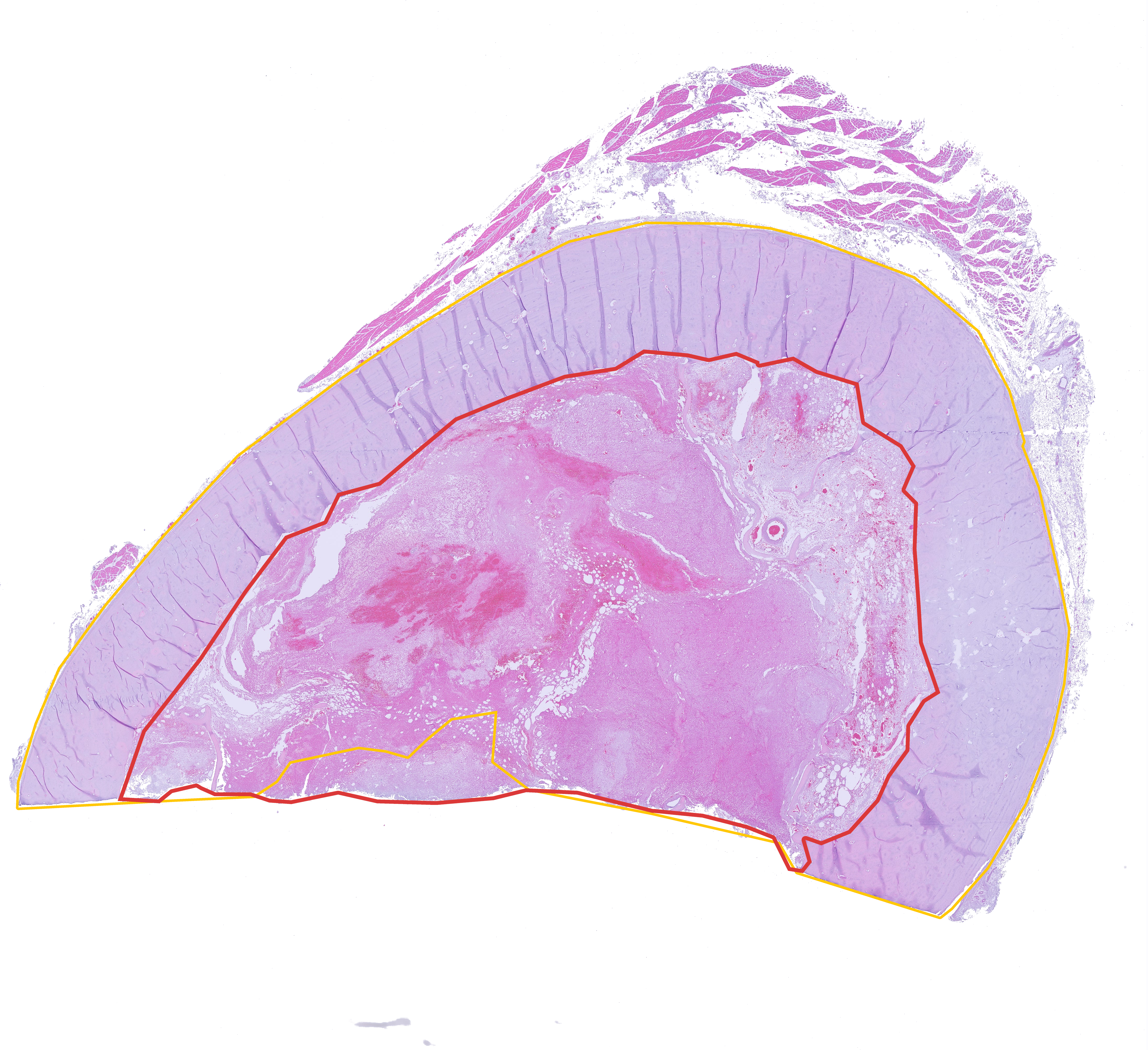}
    \caption{Image of histology slide with (yellow) bone and (red) tumour labelmaps. Image and labelmaps generated from QuPath.}
    \label{fig:Hist_Med_5}
\end{figure}

\begin{figure*}[tph]
    \centering
    \begin{subfigure}[t]{0.49\textwidth}
        \centering
        \includegraphics[height = 6cm]{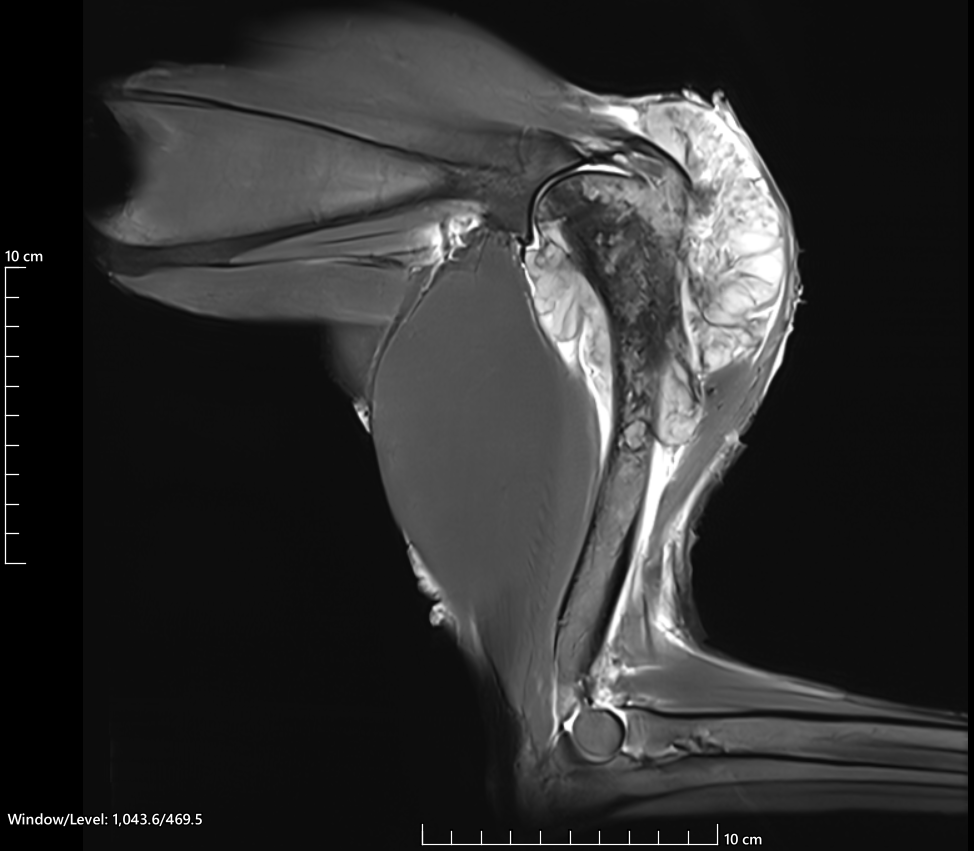}
        \caption{Sagittal cross-section of $MR_{pre}$. T2 Turbo Spin Echo sequence acquired with 3.5\,mm slice thickness.}\label{fig:ImgSub1}
    \end{subfigure}%
    \hfill
    \begin{subfigure}[t]{0.49\textwidth}
        \centering
        \includegraphics[height = 6cm]{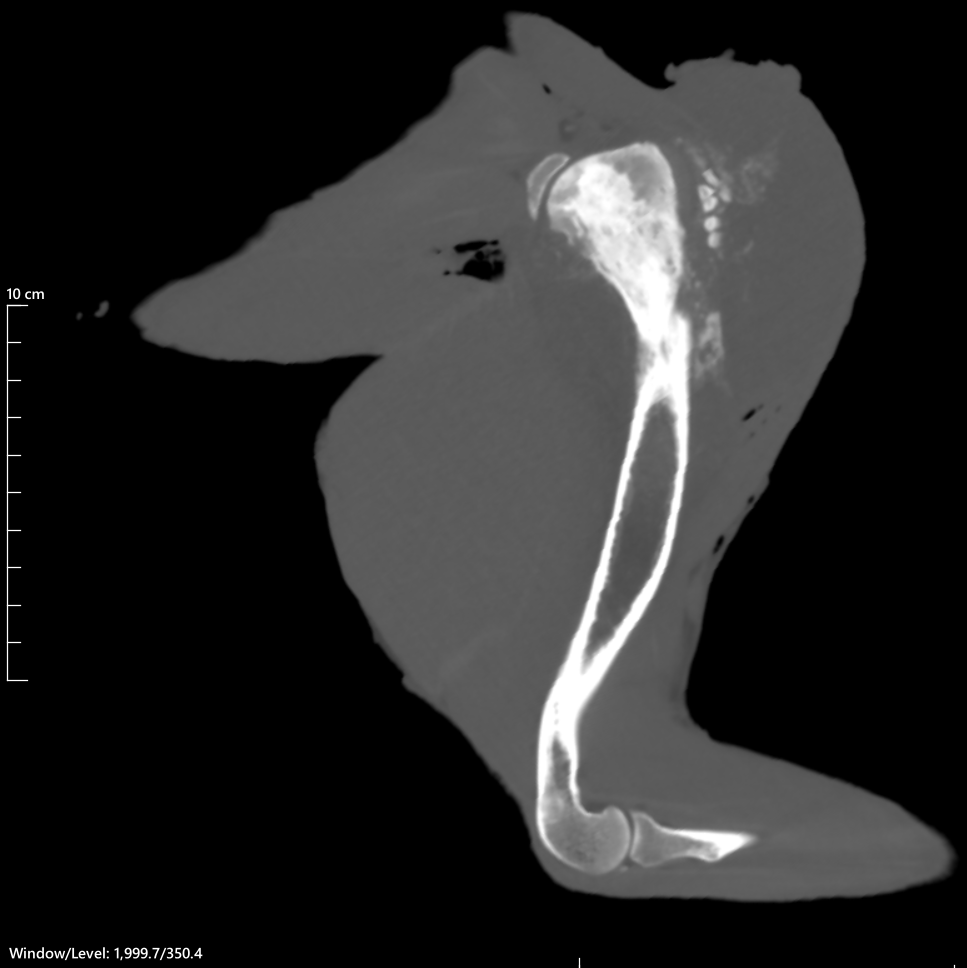}
        \caption{Reformatted cross-section of high-definition, \hbox{$CT_{pre}$} reconstruction acquired with 0.6\,mm slice thickness.}\label{fig:ImgSub2}
    \end{subfigure}
    \caption{Radiology of whole proof-of-concept canine forelimb taken on clinical equipment with pre-surgical imaging protocols. Images generated using Weasis.}\label{Fig:PrePOCRadiology}
\end{figure*}

\begin{figure*}
    \centering
    \includegraphics[width=1\linewidth]{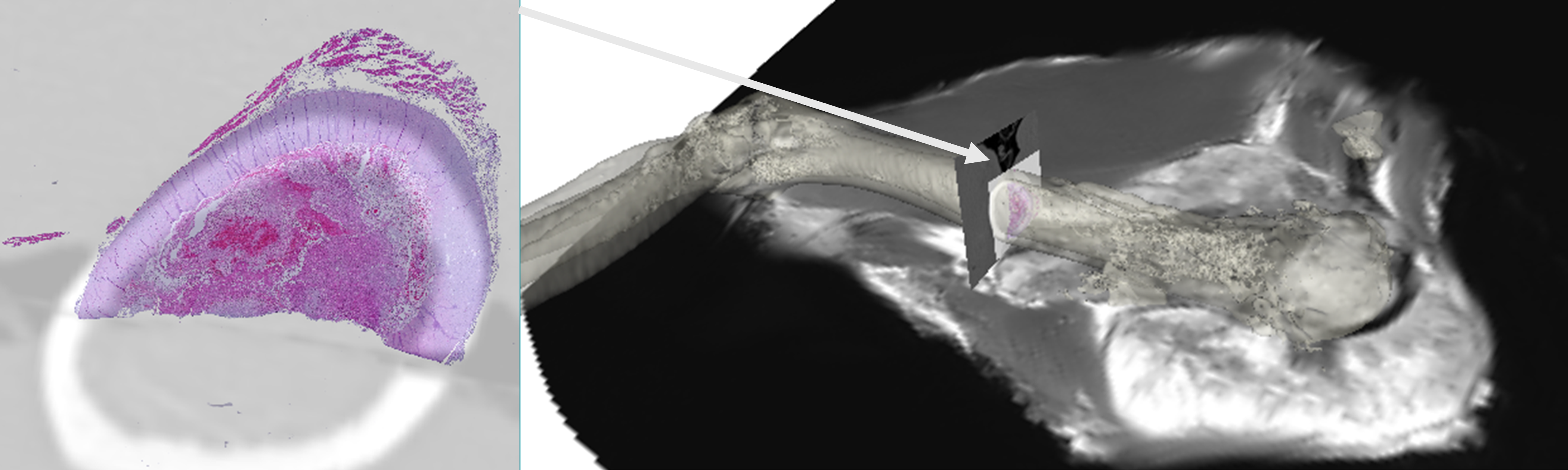}
    \caption{Image left is a single digitised axial histology image overlaid on axial $CT_{post, fused}$ cross-section. Image right is the same CT and overlaid histology orientated and overlaid on the bone segmentation of $CT_{pre}$ (opaque), along with a sagittal cross-section of $MR_{pre}$. Image generated by 3DSlicer.}\label{Fig: Overlaid histology}
\end{figure*}

Histology images and annotations (Figure~\ref{fig:Hist_Med_5}) have been co-located in two \textit{ex-vivo} CT volumes and subsequently registered to $CT_{pre}$ and $MR_{pre}$ (Figure~\ref{Fig:PrePOCRadiology}). In the proof-of-concept specimen, histology images are oriented roughly orthogonal to the humerus bone as shown in Figure~\ref{Fig: Overlaid histology}. 

Note that histology image orientation is restricted to $P_{i}$ which can take any orientation. This only depends on tissue processing decisions and measurement in pathology.

\section{Discussion}

We present a solution that enables the co-location and registration of 2D sarcoma histology images with pre-surgical CT and MR 3D volumes. This solution utilises \textit{ex-vivo} CT of resected tissue and basic laboratory measurements to restore the 3D orientation of tissue in histology, which is largely lost during current pathology processing. 


In \textbf{Step 1} the solution maintains dependence on user-guided CT bone segmentation as described in subsection \ref{Methods:StepsToSolution}. 
Fortunately, prior research has demonstrated high accuracy in CT bone segmentation, overestimating the bone boundary with a Root Mean Square (RMS) error of 0.55mm~\cite{VandenBroeck2014}.

Direct measurement of rotational or transnational plane assignment error was not feasible in the present study. Error was assessed for \textbf{Steps 1} and \textbf{2} by comparing measured dissection size against the euclidean distance between spline intersection with consecutive planes. Concerning uncertainty existed in measurement discrepancies because they didn’t initially follow a normal distribution.

This uncertainty seems to have originated, at least partially, from the assumption of flat dissection cuts, whereas pathologists recorded some as curved. The exclusion of curved dissection data supported normality (Shapiro-Wilk score of 0.367), supporting confidence in the current plane assignment validation method.

The errors in dissection cut modelling is comparable to other 3D prostate histology image registration schemes which communicated 1.1-1.9\textdegree{} and 0.9-1.3\,mm mean misalignment of histology registration~\cite{Gibson2013quantifyTissueDeformationInProstate}.

Errors between $d_2$ measurements showed higher variance than the measurements on bisection edges ($d_1$ and $d_3$), indicating rotational error in plane assignment. It is believed the perpendicular dissection assumption is weak and is worthy of further testing, especially for heavily concave bisections. 

Future work may chose to evaluate assumptions similar to how~\citeauthor{Gibson2013quantifyTissueDeformationInProstate}\cite{Gibson2013quantifyTissueDeformationInProstate} quantified registration misalignment in the prostate solution they outlined the previous year~\cite{Gibson2012prostateRegHist}. However, any revision of calculations should be carefully considered alongside the potential for histology initialisation (\textbf{Step 3}) to leverage out-of-plane histology deformation. Ongoing image processing work aims to explore the use of histology labelmaps in this.

The authors would like to note that bone tissue deformation during histology is less than that of other tissue~\cite{Cannavo2019qualEvalDeliniation}. This potentially offers a degree of stability in physical histology processing above other tissue types used in co-location work to date.

According to current clinical guidelines, initial tumour dissection should occur in the plane of the largest dimension, typically longitudinal, followed by axial sectioning~\cite{2024RCPApathologycutup}. The proof-of-concept specimen processed for this work did not undergo longitudinal dissection processing before axial cuts were made. However, the solution is flexible and can be adapted to meet specific guidelines. 

For a specimen similar in shape to the proof-of-concept, a longitudinal dissection could be processed before making axial cuts. This would provide data for a digital longitudinal histology mosaic parallel to the bisection before axial histology images are registered at an offset, orthogonal to the bisection surface.

For this work, the thickness of microtome shavings were used as a plane offset in geometric calculations. Additionally, the width of the bandsaw blade could be utilised for assigning planes if measurements are to the blockface after dissections are taken. Decisions to include these measurements will need to consider the specimen orientation during cut-up. 

A necessary level of ambiguity exists in laboratory guidelines for processing more complex specimen geometries. The ultimate responsibility of cut-up falls to pathologists~\cite{2024RCPApathologycutup}, not guidelines. Therefore maintaining a clear dialogue between staff registering digital histology, pathologists and those performing cut-up (if different personnel) is crucial to ensuring collected measurements are of value.

Laboratory technicians communicated it was relatively easy to measure dissection cuts from the tactile corner of fibrous tumour. However, it was challenging to delineate fibrous tissue from other soft tissue on CT. In this extreme proof-of-concept case (approximate tumour mass spanned 16\,cm from proximal to distal) the rotational sensitivity between planes ranged from 0\textdegree{} to 1.6\textdegree{} and plane to plane measurement showed a mean 0.08 $\pm$ 0.2\,mm discrepancy with different input markup splines. 


For this work it was assumed that a usable fiducial reference point could be created or was already present. 
A drill hole was used in processing proof-of-concept tissue however future work could use a joint condyle or the corner of bone formed by resection cuts (e.g. in chevron osteotomy). These alternatives have been shown to be identifiable on radiology and histopathology in related sarcoma work~\cite{malherbe2020assessment}.


Proof-of-concept data has been collected for demonstrating continuity of the solution. Some tissue needed reprocessing due to MR imaging equipment constraints and fixation issues, causing delays, excessive shrinkage, and lower quality in some histology images. While not suitable for characterising radiographic features, the data is available on request for understanding the work and replicating the solution.

Using animal tissue has enabled efficient prototyping and initial testing of the solution described in this work. However, the need for dedicated whole limb CT and MR with animal specimens introduced a logistical challenge. Since $CT_{pre}$ and $MR_{pre}$ are already routine in human bone sarcoma treatment workflows it is assumed this data should be available.

In this work, a custom \textit{stitch volumes} extension parameterset was made from the 5.6.2 3DSlicer release for processing the proof-of-concept CT bisecitons. This adjusted parameterset includes three modified lines and can be found in the StitchVolume file in the linked repository.

Alternatively, if customizing any parameterset is not possible, future 3DSlicer releases will offer a new \emph{StitchVolumes} extension. This extension includes a ‘Stitch Weighting Method’, which with a 0 Voxel threshold can perform \textbf{Step 4} equivalently.

No significant effort was dedicated to developing a particular extension or software package for volume-to-volume image registration because other research groups are making rapid advances in this. It is noted that pre-surgical CT and MR registration is particularly useful in complex musculoskeletal tumour resection planning~\cite{wong2008image, durr2014resection}. However, advice from higher-level associations should be used for guiding what registration software is used in sarcoma treatment centres~\cite{brock2017use,lowther2022mirsig}. 

\subsection{Development opportunities}
Some studies have reported that deep learning is a superior tool for histology co-location in certain tissues~\cite{Shao2021prostateRegHistML}. However, most existing initiatives focus on histology that has been sliced in \textit{parallel}. When tissue cannot feasibly be sliced in \textit{parallel}, as in the case of bone sarcoma, additional orientation information is required to enable co-location and registration. The use of \textit{ex-vivo} CT and laboratory measurements can provide this necessary information. The solution described in this work is freely shared as a foundation for further research to advance human sarcoma histology co-location.

While digitisation of large-format (or whole-mount) histology is rare, mosaicing techniques only require standard-size histology images. A variety of software options for this purpose are listed by \citeauthor{Pichat2018HistRecon}~\cite{Pichat2018HistRecon}. With our solution, we stress that proper labeling of dissection orientation is essential, particularly if more standard-sized histology is to be processed. Surgical dye, often used in many surgical workflows, has proven useful in maintaining specimen orientation during proof-of-concept bisection and dissection.

The assignment of $P_i$ on $CT_{post,1\&2}$ volumes and use of clinical histology labels as collected for the proof-of-concept specimen will assist histology deformation. Deformations might employ techniques such as free-form deformations \href{https://github.com/pimed/Slicer-RadPathFusion}{RAPSODI}~\cite{Rusu2020prostateRegHist}. The authors do note that non-mineralised osteoid could not be easily demarcated from
mineralised osteoid in proof-of-concept histology~\cite{Tandon2019osteoidHnEstainChallenge} due
to necessary demineralisation in pathology processing. This is expected will introduce error if deformations assume all osteoid in bone histology labelmaps is radiodense.

Single energy CT's reliance on density for tissue identification restricts it to including only mineralised osteoid in bone segmentation. Multi-energy CT may prove promising at differentiation between areas of non-mineralised osteoid and extraneous soft tissue. However, research focusing on this issue has not been found. An intermediary quantification of deformation error with any radiodense osteoid assumption would be worthy of future investigation.

\subsection{Opportunities for wider healthcare and technology}
Co-location of 2D clinical histology images within 3D radiology volumes will give physicians and researchers opportunity to contribute cohesive datasets of spatially characterised, microscopic tissue labelmaps to support radiology interpretation and surgical technology development. 
It will make high-fidelity tissue labelmaps of a clinical standard available for training segmentation algorithms at minimal additional cost. Integrating \textit{ex-vivo} bisection CT and euclidean measurements while processing proof-of-concept tissue has been shown can do this. 

Datasets of co-located histology and radiology will enhance the accuracy of volumetric segmentation by making histology insights available earlier in the bone sarcoma presentation-diagnosis-planning-surgery-recovery sequence. This will support margin decisions for surgical resection of bone sarcoma with evolving, patient-specific technologies.

\section{Conclusion}\
In this research, we present a solution to co-locate clinical bone sarcoma histology images in pre-surgical CT and MR radiology volumes. The solution is specifically engineered to convey resilience to the complexities of sarcoma histology processing. It uses laboratory measurements and \textit{ex-vivo} tumour tissue CT. A proof-of-concept bone sarcoma specimen has been processed to affirm end-to-end suitability in the clinical context.

The combination of histology and radiology through this solution will facilitate the assessment of existing treatment plan accuracy and lay the groundwork for enhanced radiology segmentation. This will support both human-managed and fully automated image volume interpretation schemes.

The computations described with this solution are capable of spatial, \textit{nonparallel} histology image co-location with an accuracy of 0.19 $\pm$ 1.8mm and user sensitivity of 0.08 $\pm$ 0.2mm. The promising misalignment error distribution suggests that continued data collection will support future efforts in quantifying histology co-location with radiology.

This work complements other histology co-location work, with possibility to build upon it~\cite{Rusu2020prostateRegHist}. It leverages existing tools in 3DSlicer and contributes a robust solution to the literature for co-location of histology images, and registration with 3D radiology.

\backmatter

\bmhead{Acknowledgements}
The authors thank Christchurch Cancer Research Trust for resourcing, veterinarians and clinics around Canterbury for logistical support and the 3DSlicer community for technical support.

\bmhead{Author Contributions}
Material preparation, data collection and analysis were performed by Keren Dittmer, Robert Phillips. Nicolas Robilliard assisted with data collection and performed histology labelmaps. The solution was developed by Robert Phillips with support from Constantine Zakkaroff. All authors contributed to the final manuscript.

\bmhead{Funding}
This work was supported by a philanthropic grant from the Canterbury Cancer Research Trust. 


\bmhead{Code Availability}
Python code for the project has been publicly shared at \href{https://github.com/phiro753/Py-OrthoRadPathFusion/}{https://github.com/phiro753/Py-OrthoRadPathFusion} 

\bmhead{Data Availability}
Proof-of-concept data is available from the corresponding author upon reasonable request.

\section*{Declarations}

\subsection*{Ethics approval}
Collection and use of animal tissue in this work did not affect animal treatment. The Otago University Ethics Committee confirmed that animal ethics was not required prior to the study commencing. 

\subsection*{Consent to participate}
Written informed consent was obtained from legal animal owners for use of tissue and data in research and education. 


\subsection*{Conflict of interests} 
The authors have no relevant financial or non-financial interests to disclose.

\bibliography{sn-bibliography}

\end{document}